\documentstyle[psfig,aps,prl,multicol]{revtex}
\begin{document}
\title{Accessible information in quantum measurement}
\author{N. J. Cerf$^{1,3}$ and C. Adami$^{1,2,3}$}
\address{$^1$W. K. Kellogg Radiation Laboratory and 
         $^2$Computation and Neural Systems\\
California Institute of Technology,
Pasadena, California 91125\\
$^3$Institute for Theoretical Physics, University of California,
Santa Barbara, California 93106}

\date{November 1996}

\draft
\maketitle
\begin{abstract}
The amount of information that can be accessed via measurement of a
quantum system prepared in different states is limited by the Kholevo
bound.  We present a simple proof of this theorem and its extension to
sequential measurements based on the properties of quantum conditional
and mutual entropies.  The proof relies on a minimal physical model of
the measurement which does not assume environmental decoherence, and
has an intuitive diagrammatic representation. 

\end{abstract}
\pacs{PACS numbers: 03.65.Bz,05.30.-d,89.70.+c
      \hfill KRL preprint MAP-207}
\vskip -0.5cm
\begin{multicols}{2}[]
\narrowtext

A fundamental issue of quantum information theory is the maximum
amount of information that can be extracted from a quantum system.
Kholevo~\cite{bib_kholevo} proved that, if a system is prepared in a
state described by one of the density operators $\rho_i$ ($i=1,\cdots
n$) with probability $p_i$, then the information $I$ (defined in the
sense of Shannon theory~\cite{bib_shannon}) that can be gathered about the
identity of the state never exceeds the so-called Kholevo bound
\begin{equation}
I \le S\left(\sum_i p_i \rho_i\right) - \sum_i p_i S(\rho_i)\;,
\end{equation}
where $S(\rho)=-{\rm Tr}\rho\log\rho$ is the von Neumann entropy.
This result holds for any measurement that can be performed on the
system, including positive-operator-valued measures (POVM's). Since
the original conjecture by Levitin~\cite{bib_levitin} (proven by
Kholevo~\cite{bib_kholevo}), work on this subject has been devoted to
obtaining more rigorous proofs of the theorem, and to derivations of
stronger upper or lower bounds on
$I$~\cite{bib_yuen,bib_jrw,bib_caves,bib_fuchs,bib_sww}. In this
Letter, we start by revisiting the derivation proposed recently by
Schumacher {\it et al.}~\cite{bib_sww} but in terms of a unified framework
of quantum information theory  which allows a more general formulation.
Ref.~\cite{bib_sww} is based on a physical model of
quantum measurement and constitutes a notable improvement
over earlier derivations which are less transparent, and that involve
maximizing mutual Shannon entropies over all possible measurements.
Here, we show that this proof can be considerably simplified by making
use of quantum conditional and mutual entropies, and the constraints
on them imposed by the unitarity of
measurement~\cite{bib_neginfo,bib_meas}.  Also, the present
formulation allows for a straightforward extension of the Kholevo
theorem to consecutive measurements.  Such a treatment clarifies
the physical content of the Kholevo theorem, which then simply states
that the {\it classical} mutual entropy (i.e. the information
acquired via measurement) is bounded from above by the {\it quantum}
mutual entropy {\em prior} to measurement. More generally, we will
show that 
\begin{equation}
H(X{\rm:}Y)\leq S(X{\rm:}Y)\;, \label{general}
\end{equation}
where $S(X{\rm:}Y)=S(X)+S(Y)-S(XY)$ is the quantum mutual entropy
between $X$ and $Y$ constructed from the density matrix $\rho_{XY}$,
while $H(X{\rm:}Y)$
is the Shannon mutual entropy~\cite{bib_ash} obtained from the
joint probability distribution $p(x,y)=\langle
x,y|\rho_{XY}|x,y\rangle$, where $|x,y\rangle$ is an arbitrary basis
in the joint Hilbert space.
The essence of the proof can be
represented by simple arithmetic on quantum Venn diagrams, by making
use of unitarity and strong subadditivity~\cite{bib_wehrl}.  In
contrast to the derivation by Schumacher {\it et al.}, no
environment-induced decoherence is needed in the physical model of
quantum measurement, while it can be added without difficulty.

\begin{figure} 
\caption{ Entropy Venn diagram for the correlated system $XQ$ {\em
before} measurement.}\label{fig_venn1}
\vskip 0.5cm
\par
\centerline{\psfig{figure=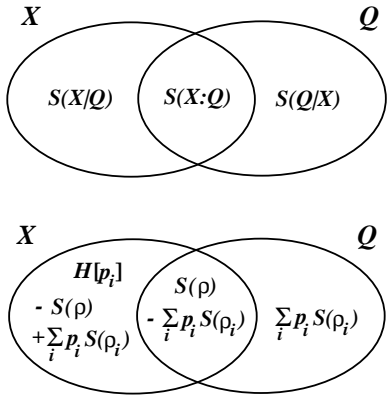,width=2.0in,angle=-90}}
\vskip 0.0cm
\par
\end{figure}

Let us assume that a ``preparer'' is described by a (discrete)
internal variable $X$, distributed according to the probability
distribution $p_i$ ($i=1,\cdots N$).  The internal state of the
preparer, considered as a physical {\em quantum} system, is then given by
the density matrix
\begin{equation}
\rho_X = \sum_i p_i |x_i\rangle \langle x_i|
\end{equation}
with the $|x_i\rangle$ being an orthonormal set of preparer states.
The state of the quantum variable $X$ can be copied to another system
simply by effecting conditional dynamics ({\it e.g.}, a controlled-NOT
quantum gate in a 2-state Hilbert space). In that sense, $X$ behaves
just like a classical variable (it can be ``cloned'') and therefore
refers to the macroscopic (collective) set of correlated internal
variables of the preparer.  Assume now that the preparer has at his
disposal a set of $N$ mixed states $\rho_i$ that can be put on a
quantum channel $Q$ according to his internal state (this is an
operation which can be performed in a unitary manner). The joint state
of the preparer and the quantum channel is then
\begin{equation}\label{decomp}
\rho_{XQ} = \sum_i p_i |x_i\rangle \langle x_i| \otimes \rho_i\;.
\end{equation}
A partial trace over $X$ simply gives the state of the quantum channel:
\begin{equation}
\rho_Q = {\rm Tr}_X \rho_{XQ} = \sum_i p_i \rho_i\equiv\rho\;.
\end{equation}
The quantum entropy of $X$ and $Q$ is $S(X) = H[p_i]$ and $S(Q)
=S(\rho)$, while
\begin{eqnarray}
S(XQ) = H[p_i] + \sum_i p_i S(\rho_i)\;,
\end{eqnarray}
where the last expression results from the fact that $\rho_{XQ}$
is block-diagonal (this is the quantum analogue of 
the ``grouping'' property of Shannon entropies~\cite{bib_ash}). The
relation between these entropies is succinctly summarized by the
quantum Venn diagram~\cite{bib_neginfo} in Fig.~\ref{fig_venn1}. 
First, let us write the quantum mutual entropy (or
mutual entanglement) between $X$ and $Q$ before the measurement (see
Fig.~\ref{fig_venn1}):
\begin{eqnarray}
S(X{\rm:}Q)&=& S(X)+S(Q)-S(XQ) \nonumber \\
&=& S(\rho)- \sum_i p_i S(\rho_i) \label{khol}
\end{eqnarray}
We see that $S(X{\rm:}Q)$ is just the Kholevo bound (the quantity 
denoted $\chi^{(Q)}$ in Ref.~\cite{bib_sww}). Thus, all we will need to prove
is that the information extracted via measurement $I\leq S(X{\rm:}Q)$.
Simple bounds for the mutual entanglement $S(X{\rm:}Q)$ can be obtained 
invoking the upper and lower bounds for the entropy of a convex combination
of density matrices (see, e.g., \cite{bib_wehrl}). Using
\begin{equation}
\sum_i p_i S(\rho_i) \le S (\rho) \le
H[p_i]+\sum_i p_i S(\rho_i)\;,
\end{equation}
implies 
\begin{equation}
0 \le S(X{\rm:}Q) \le H[p_i]\;. \label{bounds}
\end{equation}
The upper bound in (\ref{bounds}) guarantees that the entropy diagram
for $XQ$ represented in Fig.~\ref{fig_venn1} has only
non-negative entries and thus appears classical, a consequence of the fact that
$\rho_{XQ}$ was {\em constructed} as a separable state. (Negative
values for conditional entropies betray quantum non-locality, see
e.g.~\cite{bib_neginfo,bib_bell}.)

\begin{figure} 
\caption{Venn diagram summarizing the relation between entropies
(defined in the text) {\em after} measurement in the system $X'Q'A'$.}
\label{fig_venn2}
\vskip 0.5cm
\par
\centerline{\psfig{figure=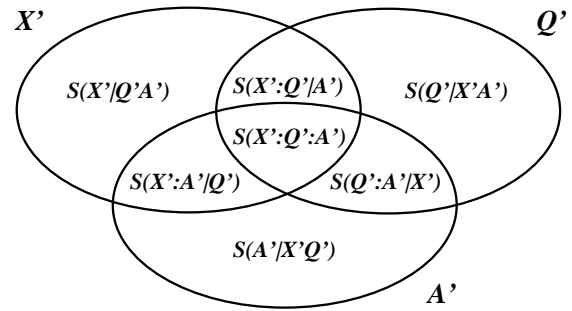,width=2.9in,angle=-90}}
\vskip 0.0cm
\par
\end{figure}

\begin{figure}
\caption{Diagrammatic representation of the Kholevo theorem using the
definitions given in Fig.~\protect\ref{fig_venn2}. Thea area enclosed
by the double solid lines represents the mutual entropy that is
conserved in the measurement $S(X'{\rm:}Q'A')=S(X{\rm:}Q)$.}
\label{fig_venn3}
\vskip 0.5cm
\par
\centerline{\psfig{figure=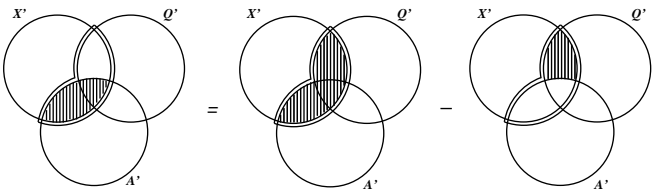,width=3.4in,angle=-90}}
\vskip 0cm
\par
\end{figure}

In the following, we describe the measurement of the preparation. This
is achieved by bringing about a unitary operation on an ancilla $A$
and the quantum preparation $Q$ that effects entanglement, and
subsequently observing the state of this ancilla. The information $I$
extracted from the measurement is then just the mutual entropy between
the ancilla and the preparer.  Before interaction, the ancilla $A$
is in a reference state $|0\rangle$ and the joint state of
the system $XQA$ is a product state $\rho_{XQ} \otimes |0\rangle
\langle 0|$. This implies of course that $S(X{\rm:}Q) = S(X{\rm:}QA)$,
as $A$ has vanishing entropy. The joint system after interaction via a
unitary transformation $U_{QA}$ is described by
\begin{equation}
\rho_{X'Q'A'} = (1_X \otimes U_{QA}) (\rho_{XQ} \otimes |0\rangle \langle 0|) 
(1_X \otimes U_{QA})^{\dagger}
\end{equation}
and the corresponding quantum Venn diagram Fig.~\ref{fig_venn2}. (We
denote by $X'$, $Q'$, and $A'$, the respective systems {\em after}
measurement.)
For the moment, let us assume that $U_{QA}$ is arbitrary; we will
discuss its specific form later. The key quantity of interest is the
mutual entanglement $S(X'{\rm:}A')$ between the physical state of the
ancilla $A$ {\it after} measurement and the physical state of the
preparer $X$ (which has remained unchanged).  We will show later that,
with certain conditions on $U_{QA}$, $S(X'{\rm:}A')$ is just the
Shannon mutual entropy between the preparer and the ancilla, or, in
other words, the information $I$ extracted by the observer about the
preparer state. Anticipating this, for a proof of Kholevo's theorem we
need only find an upper bound for $I=S(X'{\rm:}A')$.  As the
measurement involves unitary evolution of $QA$ while leaving $X$
unchanged ($X'=X$), the mutual entanglement between $X$ and $QA$ is
conserved:
\begin{equation} \label{equs1}
S(X'{\rm:}Q'A')=S(X{\rm:}QA)=S(X{\rm:}Q)\;.
\end{equation}
We now split this entropy according
to the quantum analogue of the chain rules
for mutual entropies~\cite{bib_meas,bib_channel} to obtain
\begin{equation} \label{equs2}
S(X'{\rm:}Q'A')=S(X'{\rm:}A') + S(X'{\rm:}Q'|A')
\end{equation}
where the second term on the right-hand side is a quantum conditional
mutual entropy (the mutual entropy between $X'$ and $Q'$,
conditionally on $A'$, see Fig.~\ref{fig_venn2}).  Combining
Eqs.~(\ref{equs1}) and (\ref{equs2}), we find for the mutual entropy
between ancilla and preparer after measurement 
\begin{equation}  \label{eq_egalite}
S(X'{\rm:}A')= S(X{\rm:}Q) - S(X'{\rm:}Q'|A')\;.\label{equal}
\end{equation}
This equation is represented as arithmetic on Venn diagrams in
Fig.~\ref{fig_venn3}. It indicates that the information extracted is
given by $S(X{\rm:}Q)$, the Kholevo bound Eq.~(\ref{khol}), reduced by
an amount which represents the quantum mutual entropy still existing between
the preparer's internal variable $X'$ and the quantum state after
measurement $Q'$, {\it conditional} on the observed state of the
ancilla $A'$.
The latter, the quantum conditional mutual entropy $S(X'{\rm:}Q'|A')=
S(X'A')+S(Q'A')-S(A')-S(X'Q'A')$ is in general difficult to estimate,
and we may therefore make use of strong subadditivity~\cite{bib_wehrl}
to obtain an inequality from Eq.~(\ref{equal}).  Strong subadditivity
implies that the conditional mutual entropy
$S(X{\rm:}Y|Z)=S(X{\rm:}YZ)-S(X{\rm:}Z)$ between three quantum variables
$X$, $Y$, and $Z$ is non-negative. 
This expresses the physical idea
that the mutual entanglement between $X$ and $YZ$ is larger or equal
to the mutual entanglement between $X$ and $Z$ only (just as for mutual
informations in Shannon theory), so that a mutual entanglement can
never decrease when extending a system.  
In particular we have
$S(X'{\rm:}Q'|A') \ge 0$, which implies  
$S(X'{\rm:}A')\le  S(X{\rm:}Q)$.
It remains to show that, for a particular $U_{QA}$ which describes
a measurement, the quantum mutual entropy $S(Q'{\rm:}A')$ reduces to a
Shannon mutual entropy or information, i.e., that indeed
$S(X'{\rm:}A')=I$.  Let us focus on the case of a von Neumann
measurement; using Neumark's theorem~\cite{bib_peres} it is easy to
show that the same reasoning applies to any POVM.

For the unitary evolution of a von Neumann measurement, 
we use the explicit form 
\begin{equation}
U_{QA}= \sum_{\alpha} P_{\alpha} \otimes U_{\alpha}\label{unitary}
\end{equation}
where the index $\alpha$ refers to the outcome of the measurement and
the $P_{\alpha}$'s denote projectors in the $Q$ space
associated with the measurement ($\sum_{\alpha} P_{\alpha} =1$).
The unitary operators $U_{\alpha}$ act in the $A$ space, and move
the ancilla from the initial state $|0\rangle$ to a state
$|\alpha\rangle = U_{\alpha} |0\rangle$ that points to the outcome
of the measurement. Let us assume that the $|\alpha\rangle$ are orthogonal
to make the outcomes perfectly distinguishable. The joint density matrix
after unitary evolution is given by
\begin{equation}
\rho_{X'Q'A'}= \sum_{i,\alpha,\alpha'} p_i |x_i\rangle \langle x_i|
 \otimes P_{\alpha} \rho_i P_{\alpha'}
 \otimes |\alpha\rangle \langle\alpha'| \;.
\end{equation}
Now, according to the no-collapse model of the measurement introduced
in~\cite{bib_meas}, we need to trace over the quantum system $Q'$
which induces correlations between $X'$
and $A'$. The corresponding density matrix is
\begin{equation}
\rho_{X'A'}= \sum_{i,\alpha} p_i {\rm Tr} (P_{\alpha} \rho_i)
|x_i\rangle \langle x_i| \otimes |\alpha\rangle \langle\alpha| \;.
\end{equation}
As it is a {\it diagonal} matrix, the
entropies of $X'$, $A'$, and $X'A'$ can be fully described within Shannon
theory. (Our quantum definitions of conditional and mutual
entropies reduce to the classical ones in this case~\cite{bib_neginfo}.)
A simple calculation shows that indeed
\begin{eqnarray}
S(X'{\rm:}A') &=& H \left[ {\rm Tr}(P_{\alpha}\rho) \right]
  - \sum_i p_i H \left[ {\rm Tr}(P_{\alpha}\rho_i) \right]  \\
 &=& H(A)-H(A|X) = H(X{\rm:}A)
\end{eqnarray}
where $H$ stands for the Shannon entropy and 
${\rm Tr}(P_{\alpha}\rho_i)$ is the conditional probability $p_{\alpha|i}$.
This completes our 
derivation of the standard Kholevo theorem:
\begin{equation}  \label{eq_kholevo}
I = H(X{\rm:}A)\le  S(X{\rm:}Q) = S(\rho) - \sum_i p_i S(\rho_i)\;.
\end{equation}
The present derivation emphasizes that ignoring non-diagonal matrix
elements in Eq.~(\ref{decomp}) results in a {\em classical} density
matrix (with diagonal elements $p_{i,\alpha}=p_ip_{\alpha|i}$) whose
Shannon mutual entropy is bounded from above by the corresponding {\em
quantum} entropy.  Inequality~(\ref{eq_kholevo}) arises because some
information about $X$ might still be extractable from the system $Q'$
after the measurement, i.e., $S(X'{\rm:}Q'|A')>0$ (this happens in an
``incomplete'' measurement).  This does not mean, on the other hand,
that if inequality~(\ref{eq_kholevo}) is {\em not} saturated, all of
the remaining entropy $S(X'{\rm:}Q'|A')$ can necessarily be accessed
through a subsequent measurement.  Indeed, it is known that for
non-commuting $\rho_i$'s, the bound can {\em never} be saturated, and
better upper bounds on the accessible information have been
proposed~\cite{bib_caves,bib_fuchs}.  In the picture that we have
described, inequality~(\ref{eq_kholevo}) is fundamentally linked to
the impossibility of producing a diagonal matrix $\rho_{X'Q'A'}$ with a
single $U_{QA}$ which prevents $S(X'{\rm:}Q'|A')$ from vanishing for
an ensemble of non-commuting $\rho_i$.

Note that the physical model of measurement used here~\cite{bib_meas}
is truly minimal: no environment is necessary {\it a priori}.  If
coupling to an environment $E$ {\em is} used in the description of
measurement as in~\cite{bib_sww} for diagonalizing $\rho_{X'Q'A'}$
(using decoherence as a means of selecting a ``pointer
basis''~\cite{bib_zurek}), part of the information may
flow out to this environment. In a sense, $E$ then plays the role of
another ancilla and its state after measurement can still contain some
additional information about $X$, i.e. $S(X'{\rm:}E'|Q'A')$ could be
non-vanishing.  As the environment is by definition uncontrollable,
this information can be considered to be irrecoverably lost:
$S(X'{\rm:}Q'A')\leq S(X{\rm:}Q)$. As a consequence we obtain a bound
[{\it cf.}~Eq.~(\ref{eq_egalite})]
\begin{equation}
S(X'{\rm:}A') \leq S(X{\rm:}Q) - S(X'{\rm:}Q'|A')\;.
\end{equation}
The mutual conditional entropy $S(X'{\rm:}Q'|A')$ can be calculated
explicitly in the decoherence picture using the diagonal matrix
\begin{equation}
\rho_{X'Q'A'}= \sum_{i,\alpha} p_i |x_i\rangle \langle x_i|
 \otimes P_{\alpha} \rho_i P_{\alpha}
 \otimes |\alpha\rangle \langle\alpha| \;.
\end{equation}
giving
\begin{equation}\label{schum}
S(X'{\rm:}Q'|A')=\sum_\alpha p_\alpha\left[S\left(\sum_i p_{i|\alpha}
\rho_{\alpha i}\right)-\sum_i p_{i|\alpha} S(\rho_{\alpha i})\right]
\end{equation}
where $p_\alpha={\rm Tr}(P_\alpha \rho)$, $p_{i|\alpha}=p_i
p_{\alpha|i}/p_\alpha$, and $\rho_{\alpha i}=P_\alpha \rho_i P_\alpha
/p_{\alpha |i}$ is the density matrix obtained after measuring
$\alpha$ on state $\rho_i$. (The right hand side of Eq.~(\ref{schum}) is
the quantity $\sum_\alpha p_\alpha \chi_\alpha ^{(Q)}$ of
Ref.~\cite{bib_sww}.)

Let us consider now the extension of Kholevo's theorem to many
sequential measurements. This is a generalization of the treatment of
consecutive measurements of {\em pure} states that was presented in
\cite{bib_meas}.  To that effect, let $m$ ancillae $A_1,\cdots,A_m$
interact successively with $Q$ via unitary evolutions such as
Eq.~(\ref{unitary}) with projectors $P_{\alpha_1}\cdots P_{\alpha_m}$.
The notation $A_j$ corresponds to the $j$-th
ancilla at time $j$ or later (i.e., when the $j$ first ancillae have
interacted unitarily with $Q$). As previously, unitarity implies
\begin{equation}
S(X_m {\rm:} Q_m A_1 \cdots A_m) = S(X{\rm:}Q)\;,
\end{equation}
where $X_m$ and $Q_m$ are the preparer and the quantum state after $m$
interactions. Making use of 
\begin{eqnarray}
\lefteqn{S(X_m{\rm:} Q_m A_1\cdots A_m) = 
S(X_m {\rm:}A_1 \cdots A_m)\;+}\hspace{4cm}\nonumber\\
&& S(X_m:Q_m| A_1 \cdots A_m)
\end {eqnarray}
we arrive at 
$S(X_m {\rm:}A_1 \cdots A_m)\leq S(X{\rm:}Q)$. Arguing like before, 
ignoring the non-diagonal matrix elements of $\rho_{X A_1\cdots A_m}$
yields a Shannon mutual entropy $H(X{\rm:}A_1\cdots A_m)$ based on
the conditional probabilities
\begin{equation}
p_{\alpha_1\cdots\alpha_m|i}={\rm Tr}\left(P_{\alpha_1}\cdots 
P_{\alpha_m}\rho_i P_{\alpha_m}\cdots P_{\alpha_1}\right)
\end{equation}
bounded by the corresponding quantum mutual entropy. Subsequently
using the chain relations for classical entropies,
we have the basic upper bound on the {\em sum} of accessible
informations
\begin{eqnarray}  \label{eq_multiple}
\sum_{j=1}^m H(X {\rm:} A_j| A_1 \cdots A_{j-1})
\leq S(X{\rm:}Q)  
\end{eqnarray}
where $H(X_m{\rm:}A_1|\emptyset)\equiv H(X_m{\rm:}A_1)$.
Eq.~(\ref{eq_multiple}) generalizes Kholevo's theorem and emphasizes
that the outcome of every measurement is {\em conditional} on {\em
all} previous outcomes.

Finally, inequality~(\ref{eq_kholevo}) can be shown to be a
special case of relation~(\ref{general}). Indeed, for
a general density matrix $\rho_{XY}$ describing a bipartite quantum
system whose components interact with ancillae $A$ and $B$ that define
bases $|x\rangle$ and $|y\rangle$ respectively, we have
$S(A'{\rm:}B')=H(X{\rm:}Y)$, the Shannon mutual entropy of the joint
probability $p_{xy}=\langle x,y|\rho_{XY}|x,y\rangle$.  Using
\begin{eqnarray}
S(X{\rm:}Y)&=&S(X'A'{\rm:}Y'B')\nonumber\\
&=&S(A'{\rm:}B')+S(A'{\rm:}Y'|B')+S(X'{\rm:}Y'B'|A')
\end{eqnarray}
and the non-negativity of conditional mutual entropies yields
$H(X{\rm:}Y)\leq S(X{\rm:}Y)$.

\acknowledgements We wish to acknowledge C. Caves, C. Fuchs, and
A. Peres for very useful discussions.
This research was supported in part by the
NSF under Grant Nos. PHY 94-12818, PHY94-20470 at Caltech and PHY
94-07194 at the ITP at Santa Barbara.

\end{multicols}
\end{document}